\documentclass{article}

\usepackage[utf8]{inputenc}
\usepackage{amsmath,amsfonts}
\usepackage{listings}
\usepackage{bm}
\usepackage{todonotes}
\usepackage{hyperref}

\usepackage{graphicx}
\usepackage{subcaption}

\author{Jesper L. Hinrich, Søren F. V. Nielsen, Nicolai A. B. Riis, \\ Casper T. Eriksen, Jacob Frøsig, Marco D. F. Kristensen,  Mikkel N. Schmidt, \\ Kristoffer H. Madsen and Morten Mørup \footnote{This work was supported by the Lundbeck Foundation, grant no. R105-9813.} \\
Technical University of Denmark}

\title{Scalable Group Level Probabilistic Sparse Factor Analysis}

\def\btheta{\boldsymbol{\theta}}
\def\bmu{\boldsymbol{\mu}}

\def\<{\langle}
\def\>{\rangle}
\def\bgamma{\boldsymbol{\gamma}}
\def\balpha{\boldsymbol{\alpha}}

\def\btau{\boldsymbol{\tau}}

\def\A{{\bf A}}
\def\a{{\bf a}}

\def\I{{\bf I}}

\def\S{{\bf S}}
\def\s{{\bf s}}

\def\X{{\bf X}}
\def\x{{\bf x}}

\def\0{{\bf 0}}
\def\1{{\bf 1}}

\def\<{\, \left\langle \,}
\def\>{\, \right\rangle \,}

\def\tr{\mathrm{trace}}
\def\diag{\mathrm{diag}}

\def\balpha{\boldsymbol{\alpha}}

\def\bSigma{\boldsymbol{\Sigma}}

\def\btau{\boldsymbol{\tau}}

\def\btheta{\boldsymbol{\theta}}

\def\DN{\mathcal{N}}

\def\Uni{\mathrm{Uniform}}

\def\DGamma{{\cal G}}

\def\ctilde{\kern -.04em\lower .7ex\hbox{\~{}}\kern .04em}

\usepackage{placeins} 

\begin{document}

\maketitle

\begin{abstract}
Many data-driven approaches exist to extract neural representations of functional magnetic resonance imaging (fMRI) data, 
but most of them lack a proper probabilistic formulation.
We propose a group level scalable probabilistic sparse factor analysis (psFA) allowing spatially sparse maps, component pruning using automatic relevance determination (ARD) and  subject specific heteroscedastic spatial noise modeling.
For task-based and resting state fMRI, we show that the sparsity constraint gives rise to components similar to those obtained by group independent component analysis. The noise modeling shows that noise is reduced in areas typically associated with activation by the experimental design. 
The psFA model identifies sparse components and the probabilistic setting provides a natural way to handle parameter uncertainties. The variational Bayesian framework easily extends to more complex noise models than the presently considered.
\end{abstract}

\vspace{-.4cm}\section{Introduction}
In functional magnetic resonance imaging (fMRI) large amounts of data are currently being generated due to high spatial resolution, an increase in the typical number of session acquired and a trend towards multi-center acquisition and data sharing. There is therefore a growing need for methods and algorithms that scale, while still keeping reasonable model assumptions. 

 
A common problem in functional neuroimaging is finding a good latent representation of the data.
One approach is to use data-driven methods, which allows the data to ``speak for itself''. Two popular approaches in fMRI is to assume orthogonality or independence, giving rise to principal component analysis (PCA) \cite{andersen1999principal} and independent component analysis (ICA) \cite{mckeown1997analysis}, respectively. For group level analysis group-ICA \cite{Calhoun2001-lt} and independent vector analysis (IVA) \cite{Adali2014-tu} are the most prominent data-driven methods for inferring components of neural activity. Group-ICA and IVA have the advantage that independent components often are sparse providing interpretable spatial activation maps, due to the typical assumption of sparse source distributions.. This interpret-ability in fMRI may also be achieved by models optimizing for sparsity instead of independence \cite{daubechies2009independent}, which in some cases achieves similar results despite optimizing different properties \cite{calhoun2013independent}.

ICA models in general assume noise free data, which is commonly achieved by modeling additive homoscedastic Gaussian noise, or the noise is sought removed through a combination of (PCA) pre-whitening and standardization to unit voxel variance. A probabilistic ICA (pICA) approach with spatial and temporal noise modeling was suggested in \cite{beckmann2004probabilistic}, which included a noise estimation step in their framework. However, the noise is estimated in a separate step from the ICA components, which can be problematic in comparison with joint estimation \cite{chen1993joint}. While \cite{beckmann2004probabilistic} does incorporate probabilistic elements into ICA, it is not probabilistic from a Bayesian modeling perspective.  In contrast, joint estimation has shown promising results on fMRI (cf. \cite{Roge2015-jo,Hinrich2016-dd}).

%
%
We propose a probabilistic sparse factor analysis (psFA) model for group level analysis of fMRI data with heteroscedastic noise. Model inference is done using variational Bayes with a mean-field approximation and automatic relevance determination (ARD) \cite{MacKay1995-gg} to promote sparsity on individual voxels of the spatial maps and prune components by learning their relevance in time.
To overcome the large computational burden involved 
massive parallelization of the updates is exploited using a graphical processing unit (GPU). The model is first investigated on synthetic data, where the advantages of using a Bayesian approach to factor analysis (FA) and principal component analysis (PCA) is briefly assessed. The proposed model's applicability to fMRI is tested on a motor task experiment \cite{rasmussen2012model} and on a resting state experiment \cite{Poldrack2015-fk} and contrasted with pICA (MELODIC\cite{beckmann2004probabilistic}).


\section{Methods and Data}
\subsection{Probabilistic Sparse Factor Analysis}
We propose a group level probabilistic sparse factor analysis model (psFA), which is a combination of the probabilistic sparse PCA model proposed in \cite{guan2009sparse}, the group level PCA analysis proposed in \cite{hinrich2016grouppca} and the inclusion of heteroscedastic voxel noise, first proposed in the context of fMRI and variational inference in \cite{nielsen2004variational}. The generative model for a data array, $\X \in \mathbb{R}^{V \times T \times B}$, with $V$ voxels, $T$ timepoints and $B$ subjects, can be written as,

\begin{align*}
\alpha_{vd} \sim \DGamma(a_\alpha,b_{\alpha_{vd}}), \gamma_d \sim \DGamma & \left(a_\gamma,b_{\gamma_{d}}\right), \tau_v^{(b)} \sim \DGamma \left(a_\tau,b_{\tau_{v}^{(b)}}\right) \\
\a_v \sim \DN \left(\0,\diag\left(\balpha_{v}\right)^{-1}\right) , &\; \s_t^{(b)} \sim \DN\left(\0,\diag(\bgamma)^{-1}\right) \\
\x_t^{(b)} \sim \DN(\A\s_t^{(b)},&\;\diag(\btau^{(b)})^{-1}),
\end{align*}
in which $d$ indexes the latent space dimension $D$, $v$ indexes voxels, $t$ indexes time and $b$ indexes subjects. The parameter $\alpha_{vd}$ is the precision on the spatial maps in the matrix $\A$, and acts as a sparsity pattern. The parameter $\gamma_d$ is the precision on the $d$'th component in the time-courses. Thus we have two 'forces' that can prune in the model, $\alpha$ to make the maps sparse and $\gamma$ to prune away irrelevant components.

Finding the posterior $P(\btheta|\X)$ is analytically intractable and an approximate solution is found through variational inference, as originally proposed for PCA by \cite{Bishop1999-si}. For the psFA model the mean-field approximation to the posterior,
\begin{align*}
Q(\btheta|\X) = \prod^{V}_{v=1} \DN(\a_v|\bmu_{\A_v},\bSigma_\A^{(v)}) \prod^{B,T}_{b,t=1}\DN(\s_t^{(b)}|\bmu_{\S_t}^{(b)},\bSigma_{\S}^{(b)})\\ \prod^{B,V}_{b,v=1}\DGamma(\tau_{v}^{(b)}|\tilde{a}_\tau,\tilde{b}_{\tau_{v}^{(b)}}) \prod^{D}_{d=1}\DGamma(\gamma_{d}|\tilde{a}_\gamma,\tilde{b}_{\gamma_{d}}) \prod^{D,v}_{d,v=1}\DGamma(\alpha_{vd}|\tilde{a}_\alpha,\tilde{b}_{\alpha_{vd}}),
\end{align*} 
is used due to its similarity to the actual $P(\cdot)$ distributions yielding closed form solutions in the update rules. We use coordinate ascent variational inference, updating the moments of each variational distribution in a cyclic fashion conditioning on the other moments. 
The derived moments and a MATLAB implementation of the method are provided online\footnote{\url{https://brainconnectivity.compute.dtu.dk/}}.

The computational burden of the proposed model lies in calculating $\bSigma_\A^{(v)}$ for each voxel $v$, which has $O(D^3 V)$ time complexity due to inversion of $V$ matrices of size $D\times D$. These inversions are  embarrassingly parallel, and can be calculated quickly using GPUs, but comes at the price of having to keep the matrices in memory, requiring $O(D^2 V)$ space.

If desired, subject specific mean values can be modeled such that $\X^{(b)}=\A\s_t^{(b)}+\bmu^{(b)}$,  where $\bmu^{(b)}\sim \DN(\0,\beta^{-1}\I_V)$. In practice, we have removed the empirical mean values prior to analysis. 

\subsection{Motor-task Data}
\label{sec:method_motor}
We investigate the proposed model on a motor task experiment, which was previously acquired  and analyzed in \cite{rasmussen2012model,rasmussen2012nonlinear}. The experiment consisted of $B=29$ young and healthy adults, scanned while performing a block design motor task. The participants were visually cued by a blinking light, to indicate either right (green light) or left (red light) hand finger tapping. Each scanning session consisted of 10 task blocks, where each block consisted of a sequence of four tasks, i.e. ``right/rest/left/rest'', with in total 240 images for each session. 
The data was pre-processed using standard techniques and parameter settings of the SPM8 software package\footnote{\url{http://fil.ion.ucl.ac.uk/spm}}. Each subject was realigned to the mean volume (rigid-body), normalized to a common Montreal Neurological Institute (MNI) template, and resliced to native 3 $\text{mm}^3$ resolution. Afterwards spatial smoothing with a 3D Gaussian kernel (6mm FWHM) was applied, wavelet despiking to remove temporal outliers \cite{patel2014wavelet}, voxel means subtracted, and data detrended via high-pass filtering with a 128 s cutoff. Finally we applied a rough grey-matter mask with 48799 voxels, and afterwards each subject was z-scored individually.

\subsection{Resting State Data}
We used the resting-state data\footnote{\url{https://openfmri.org/dataset/ds000031/}} from \cite{Poldrack2015-fk} and applied the following pre-processing steps to sessions 014-104\footnote{Some sessions did not contain resting state data and were thus discarded} using SPM12. All sessions were coregistred to the first image of the first functional session (session 014), and all sessions were then jointly corrected for motion artefacts using a rigid-body transformation towards the mean volume. A T1-weighted anatomical image from session 012 was coregistred to the functional space and grey matter (GM), white matter (WM) and cerebrospinal fluid (CSF) was segmented using the standard tissue probability map from SPM. All functional sessions were then highpass filtered (1/128 Hz), nuisance regressed using motion parameters and eroded CSF and WM masks, and wavelet despiked \cite{patel2014wavelet}. Finally, all sessions were resliced (due to a change in the number of slices after session 027) to the first session and smoothed using a FWHM 5mm Gaussian kernel. The GM-mask was then resliced to the functional images and thresholded yielding a data matrix of size 69430 voxels $\times$ 518 timepoints for each session. Due to memory limitations on the GPU we only considered the 25 first sessions of the data. For visualization purposes, we normalized the components from psFA and MELODIC to MNI space (2 mm$^3$ resolution) using the deformation field estimated in the segmentation step.

\section{Results}
For all analysis the psFA (or pFA) model  the following parameters are fixed; $a_\gamma$, $a_\alpha$, $a_\tau$, $b_{\gamma_d}$, $b_{\tau_{v}^{(b)}}$, $b_{\alpha_{v,d}} = 1\mathrm{e-}6$. Variational inference was performed for the remaining parameters (except $\balpha$ for pFA), starting form an initial initial solution where the elements of $\A$ were drawn from a $\DN(0,1)$ distribution and the subject specific time courses were then back reconstructed,  $\S^{(b)} = (\A^T \A)^{-1} \A^T \X^{(b)} $ . In our analysis we try to mitigate the effect of local minima by running the psFA analysis multiple times with random initializations. We note that this is not sufficient to avoid local minima, and this should be investigated further. In all results in this section, only the run achieving the maximum lowerbound is further analysed. 

\subsection{Synthetic Experiments}
We investigate the model in a synthetic setting with $B=3$ subjects. For each subject, we generated three sources ($D=3$), of length $T = 25$, from a normal distribution with zero mean and unit variance. These sources were then mapped to a higher dimensional space of size $V = 1000$ through a sparse matrix $\A$. The elements of $\A$ were generated from a $\DN(0,1)$ and element-wise multiplied by a binary indicator from $\Uni(0,1)>0.5$. Heteroscedastic voxel- and subject-specific noise variance $\tau_v^{\vspace{-0.2cm}(b)^{-1}}$  was drawn from $\DN(0.009,0.002)$. 
Noise drawn from $\DN(0,\tau_v^{\vspace{-0.2cm}(b)^{-1}})$ was then added to the corresponding voxel and subject, yielding a data set of size $V \times T \times B$. 

\begin{figure}[t]
 \centering
 \begin{tabular}{ccccc}
 \textbf{True $\A$} &\textbf{psFA}	& \textbf{pFA} & \textbf{ICA} & \textbf{PCA}\\
	\hline \hline \vspace{0.5cm}
    {\hspace{-0.3cm} \includegraphics[width = .17\linewidth]{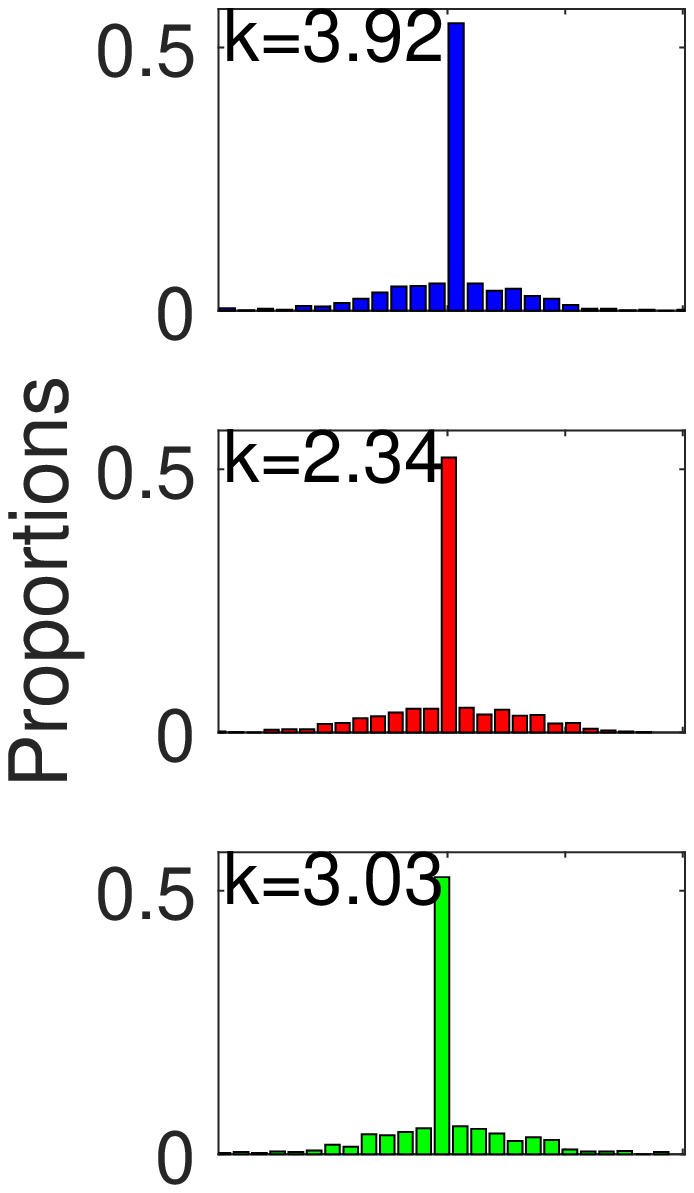} } 
		&{\hspace{-0.3cm}\includegraphics[width = .17\linewidth]{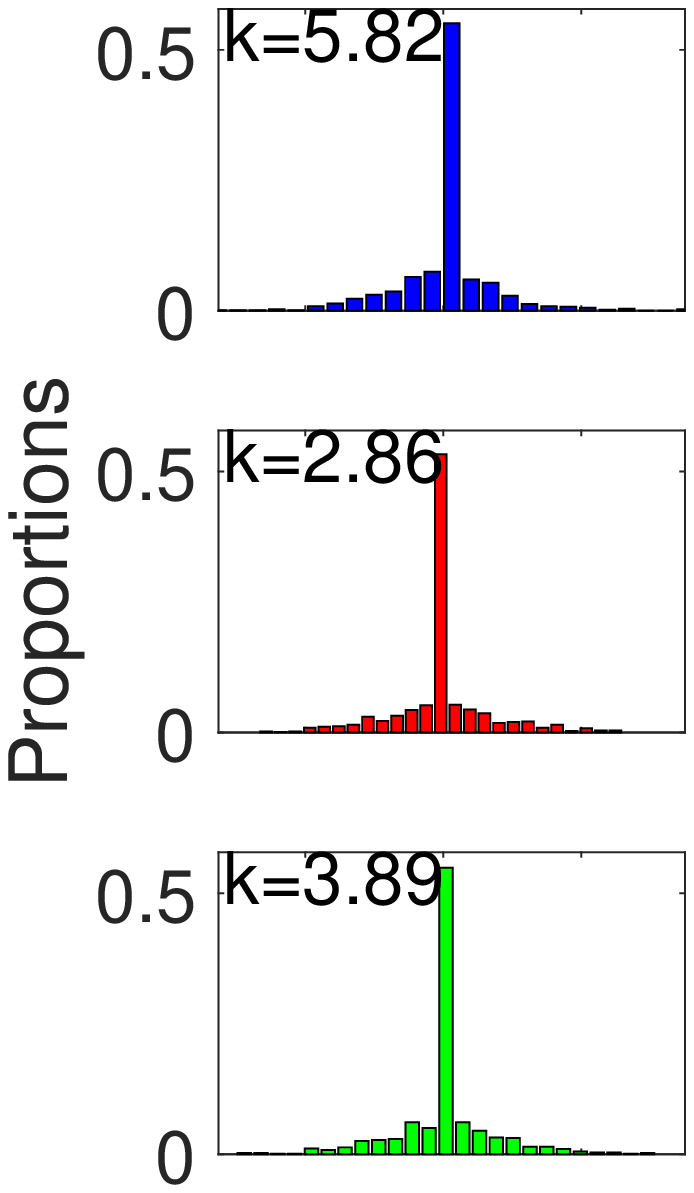} }
        &{\hspace{-0.3cm}\includegraphics[width = .17\linewidth]{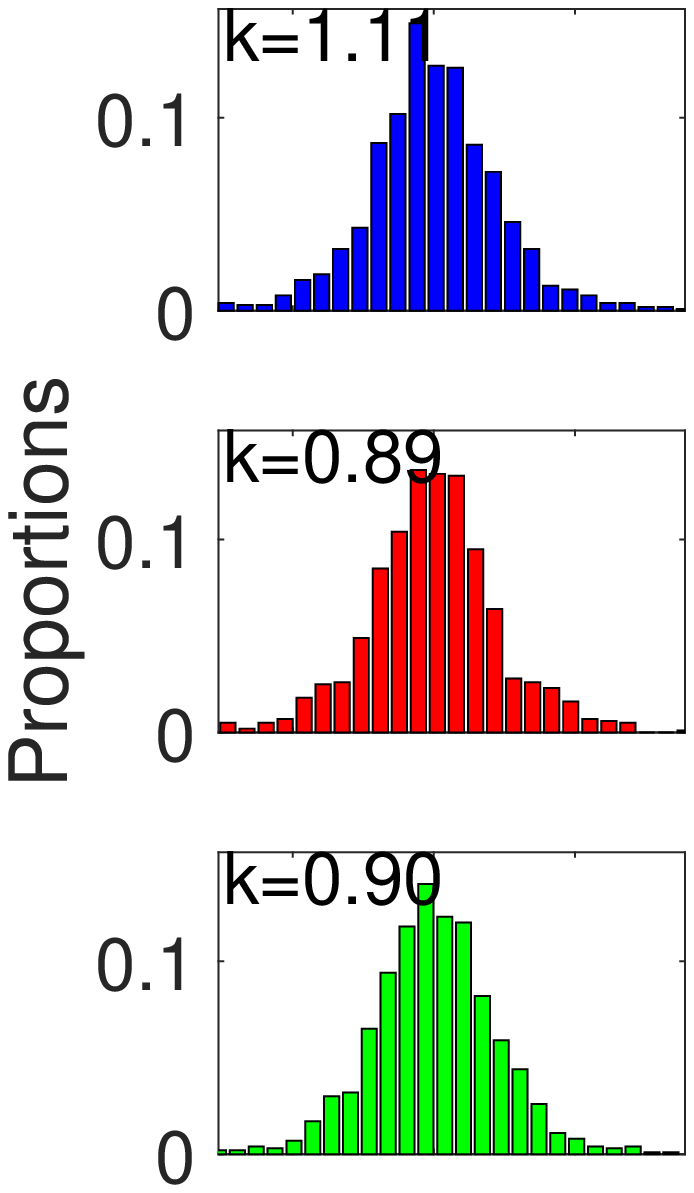} }
        &{\hspace{-0.3cm}\includegraphics[width = .17\linewidth]{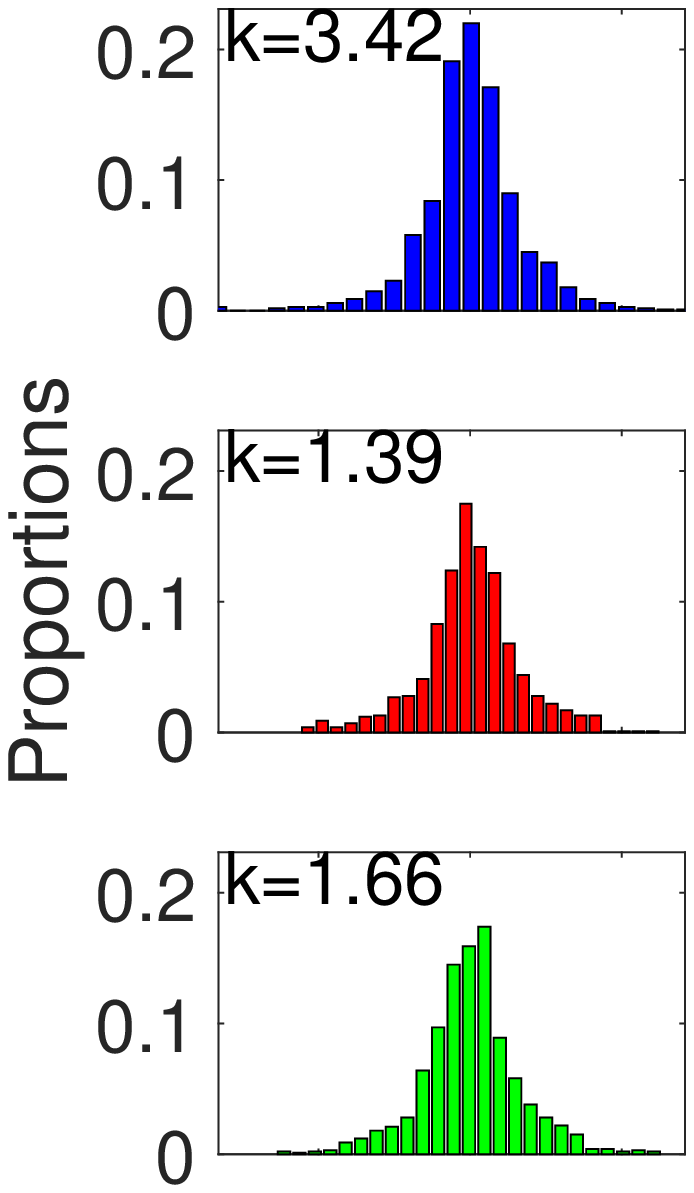} }
        &{\hspace{-0.3cm}\includegraphics[width = .17\linewidth]{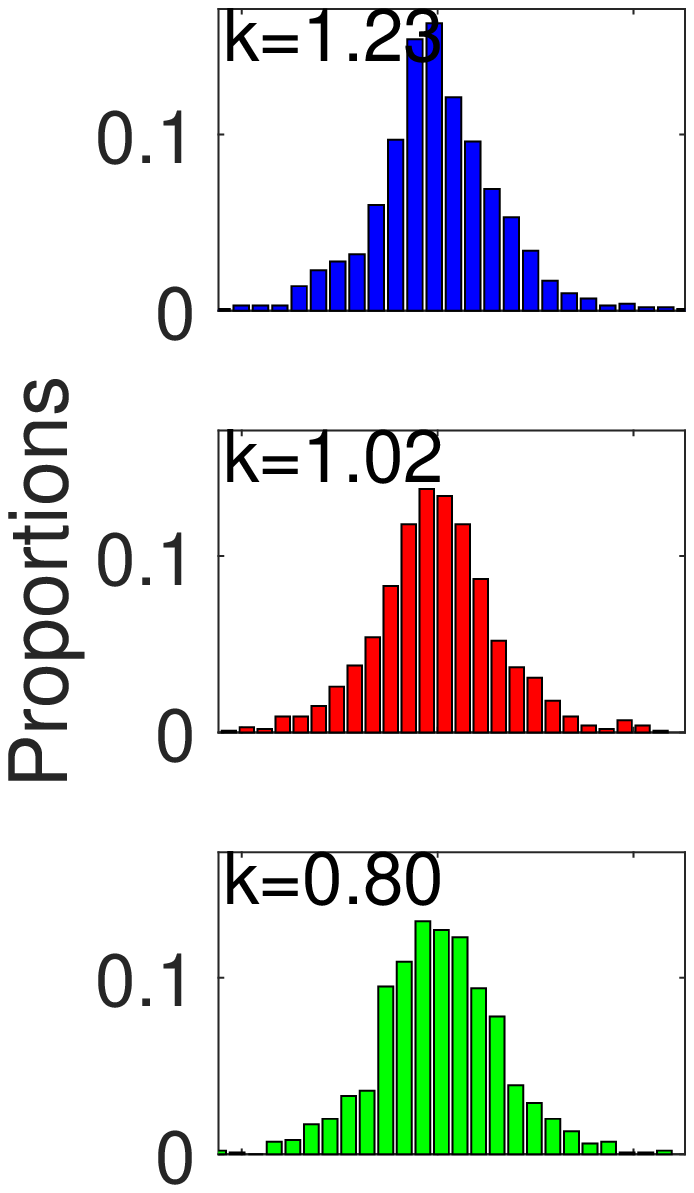} } \vspace{-0.5cm}\\ 
    Amari: &$0.097$ & $2.005$ & $0.090$  & $1.641$\\
    Corr.: &$0.881$& $0.694$& $0.825$ & $0.753$\\
    \hline & & & &
 \end{tabular}
 \vspace{-0.75cm}\caption{\textbf{Synthetic experiments}: Comparison between the true and estimated (first three) spatial maps $\A$. A histogram for each map is shown. Further, the Amari distance and average correlation between $\A_{\mathrm{true}}$ and $\A_{\mathrm{est}}$ is given.}\label{fig:synthetic}
\vspace{-.3cm}
\end{figure}

The psFA (sparsity) and pFA (no sparsity) model were then run for 500 iterations with $D=6$, with fifty random restarts, results are shown in Fig.~\ref{fig:synthetic}. These methods are compared to regular PCA and infomax ICA\footnote{\url{http://cogsys.imm.dtu.dk/toolbox/ica/}}  on temporally concatenated data. The best performing methods are psFA and ICA, achieving high correlation and low Amari distance \cite{bach2002kernel}.


\subsection{Motor-task Experiment}
\label{sec:results_motor}
The psFA and pFA models were run for 1000 iterations with $D=25$ components and we restarted the algorithm 5 times. The results obtained by psFA and pFA are contrasted to those found by MELODIC-ICA \footnote{\url{http://fsl.fmrib.ox.ac.uk/fsl/fslwiki/MELODIC}}  (pICA) with default settings, but using the same grey matter mask as described in section~\ref{sec:method_motor}. 

\begin{figure*}
\centering
\begin{subfigure}[b]{\linewidth}
\centering
\begin{tabular}{ccc}
Left sensorimotor related & Right sensorimotor related &  Visual related match\\
\includegraphics[width=.33\linewidth]{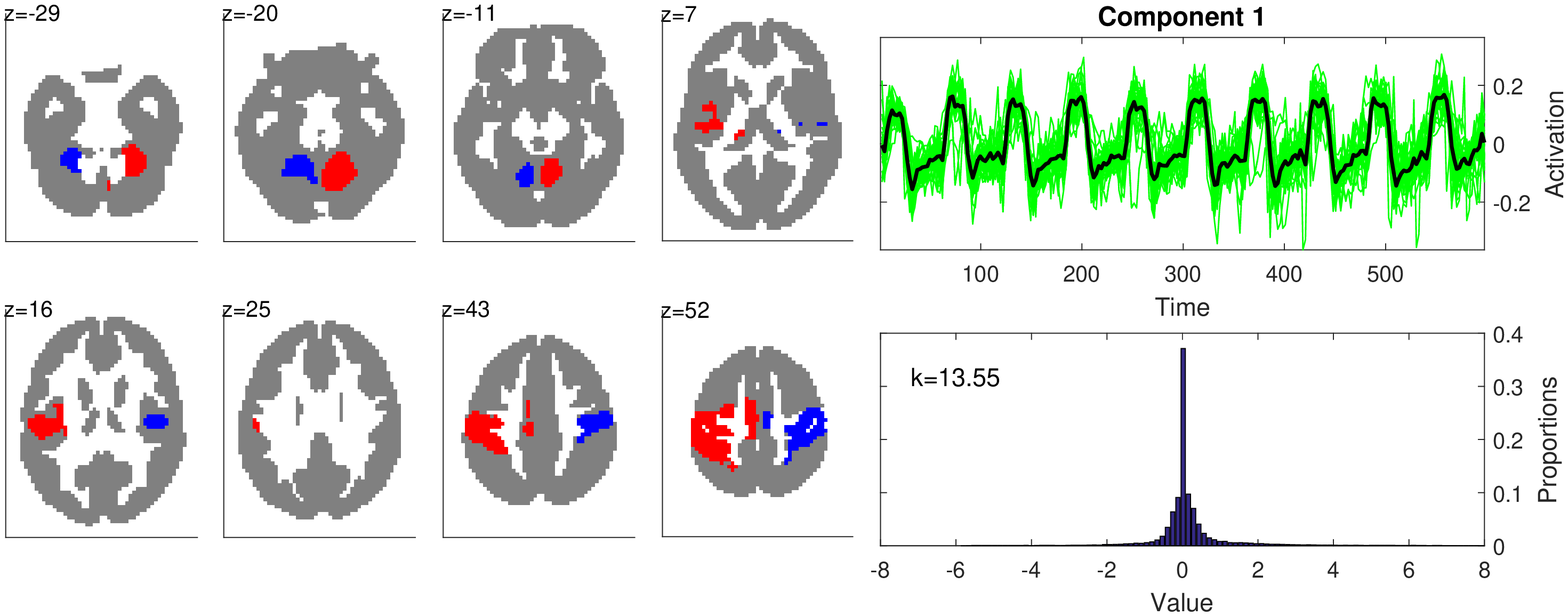} &
\includegraphics[width=.33\linewidth]{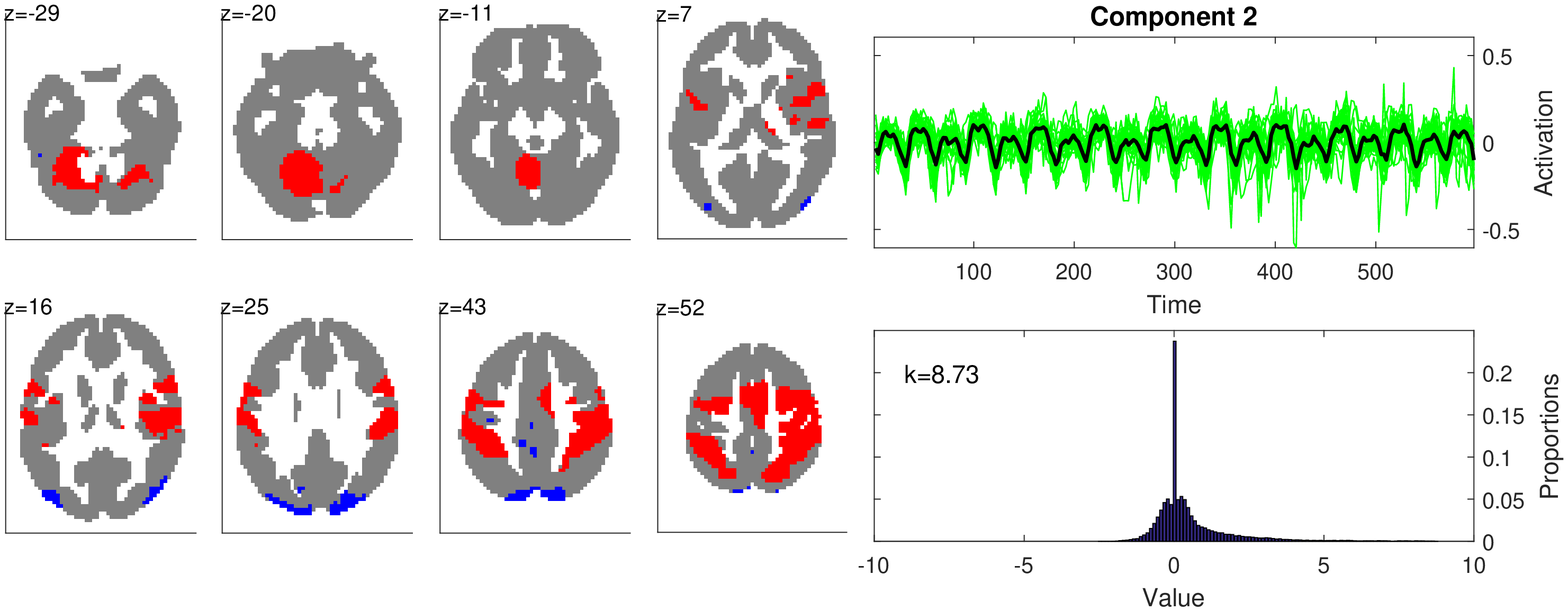} &
\includegraphics[width=.33\linewidth]{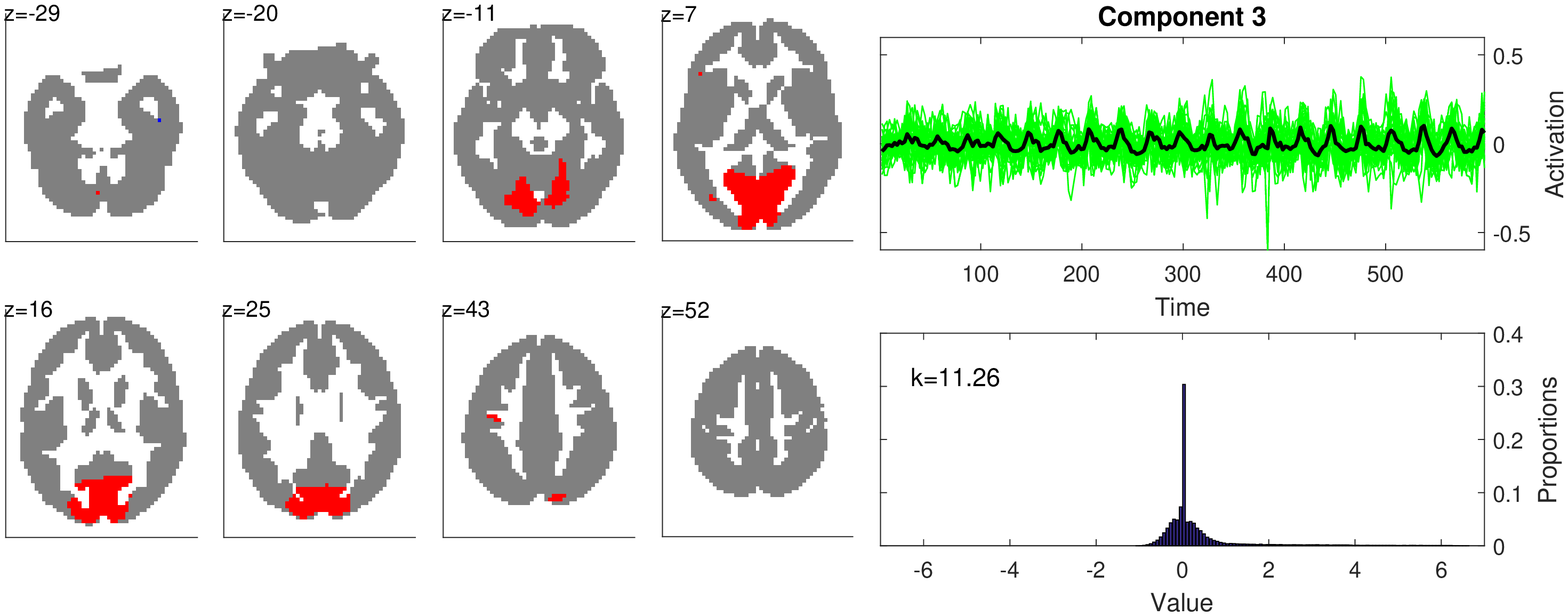}
\end{tabular}
\vspace{-.3cm}\caption{\textbf{psFA}: The estimated components shows the sparsity constraint leads to components which are more sparsely described and the histogram shows many near zero values.} \label{fig:motor_sfa}
\end{subfigure}
\begin{subfigure}[b]{\linewidth}
\centering
\begin{tabular}{ccc}
\includegraphics[width=.33\linewidth]{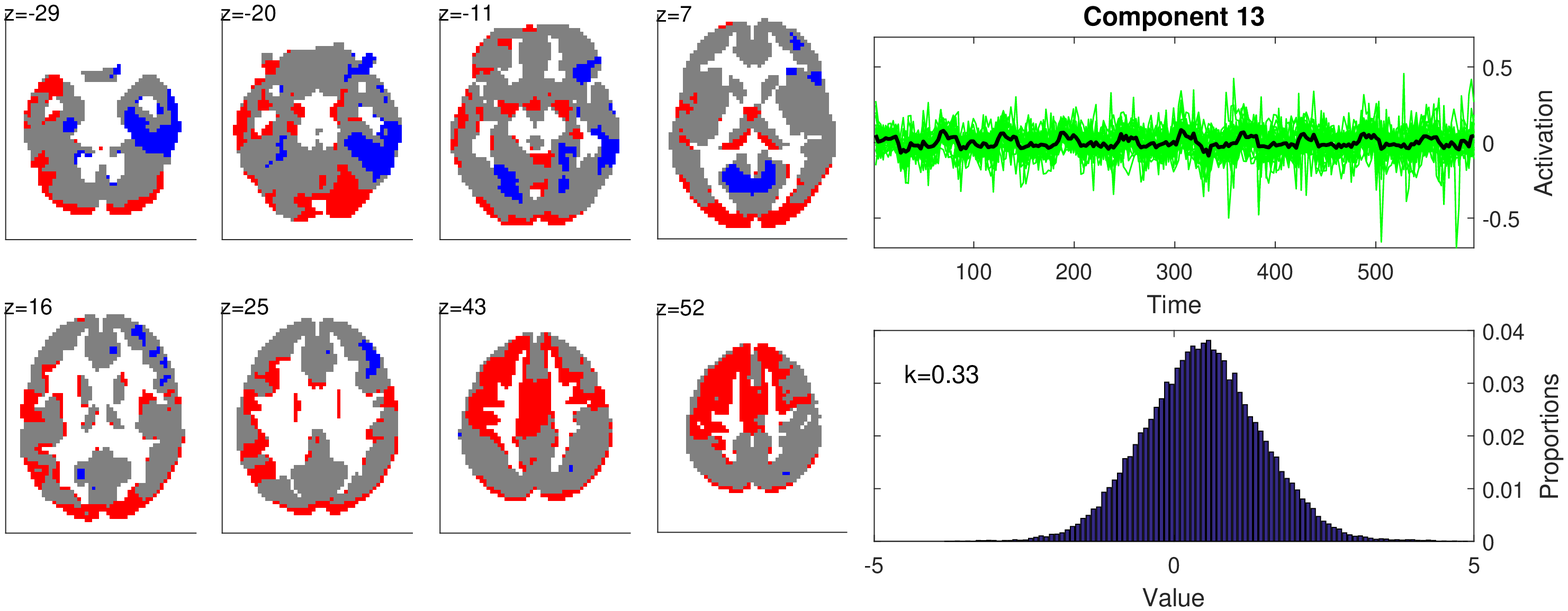} &
\includegraphics[width=.33\linewidth]{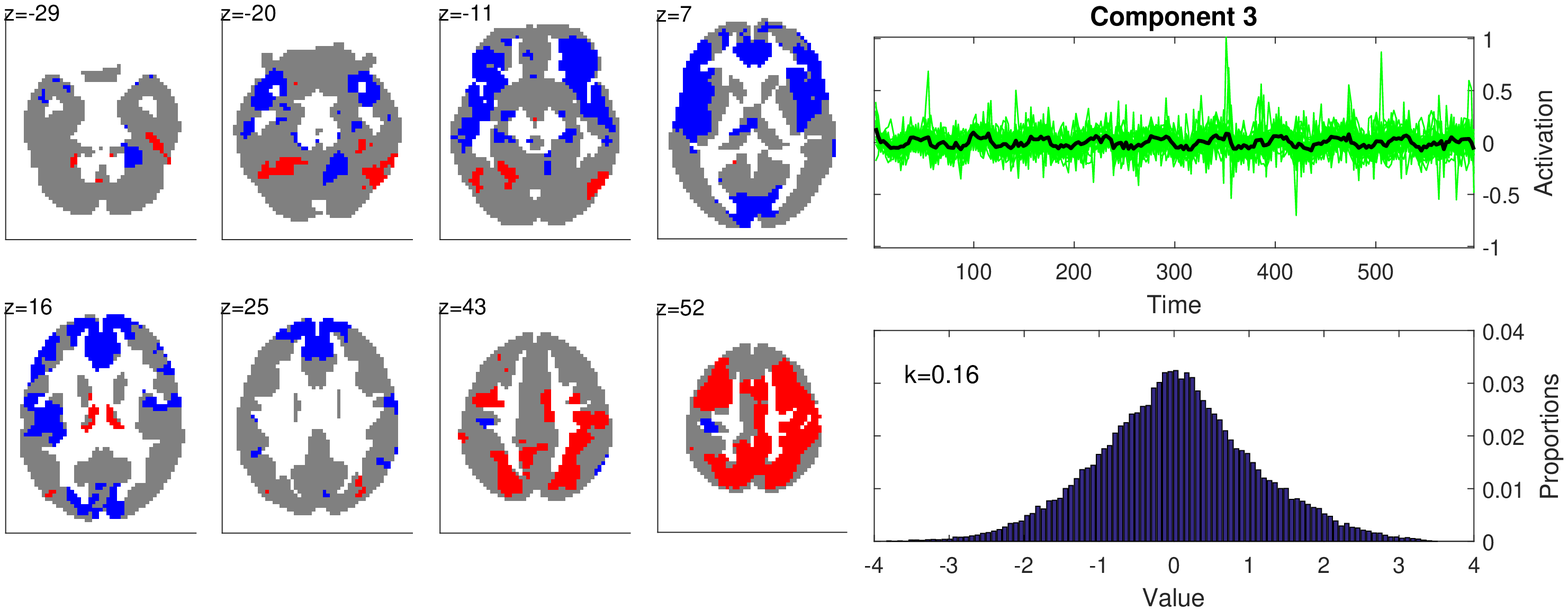} &
\includegraphics[width=.33\linewidth]{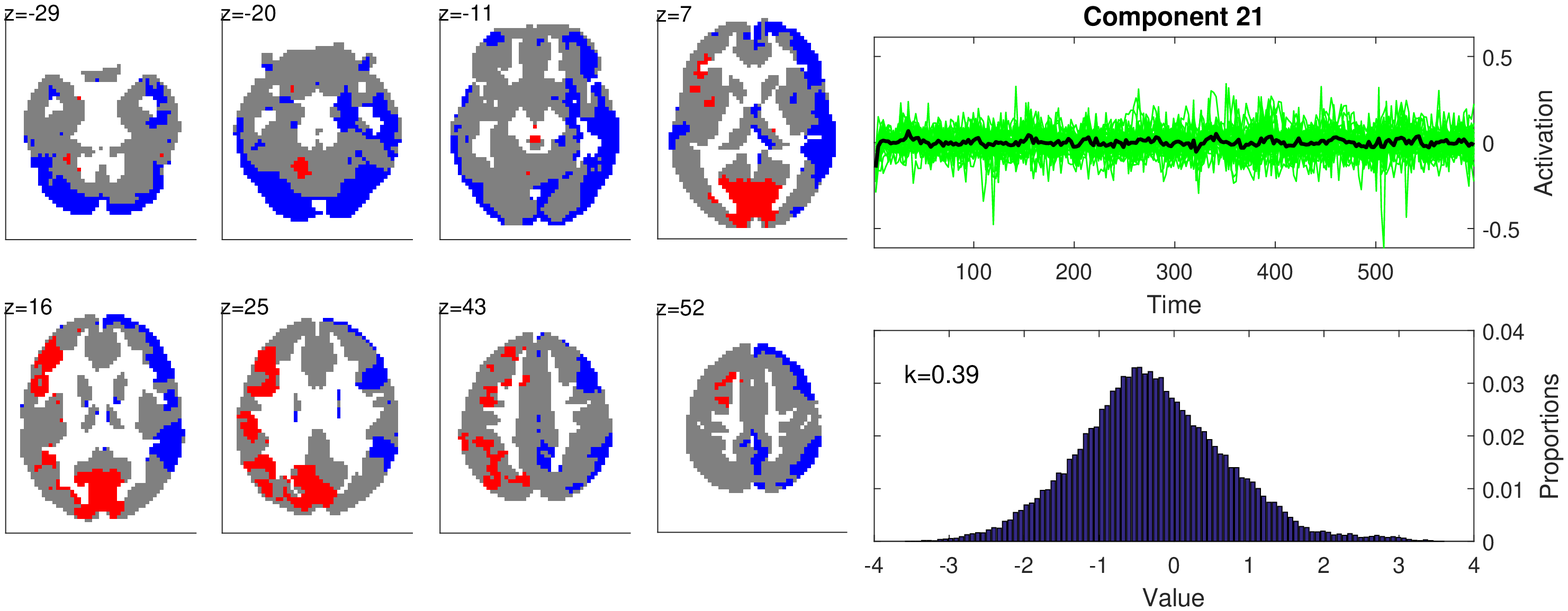}
\end{tabular}
\vspace{-.3cm}\caption{\textbf{pFA}: The model finds broad components, as can be seen from both the slices and histograms, where many voxels have high values.}\label{fig:motor_fa}
\end{subfigure}
\begin{subfigure}[b]{\linewidth}
\centering
\begin{tabular}{ccc}
\includegraphics[width=.33\linewidth]{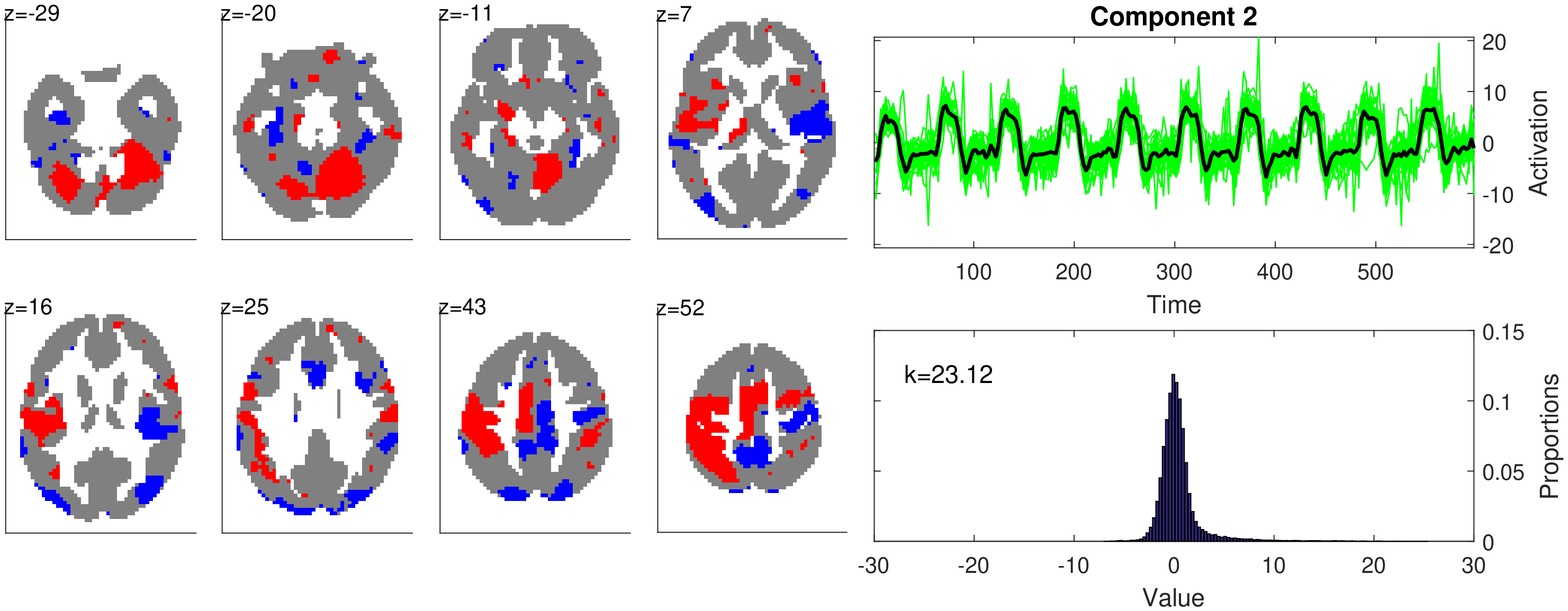} &
\includegraphics[width=.33\linewidth]{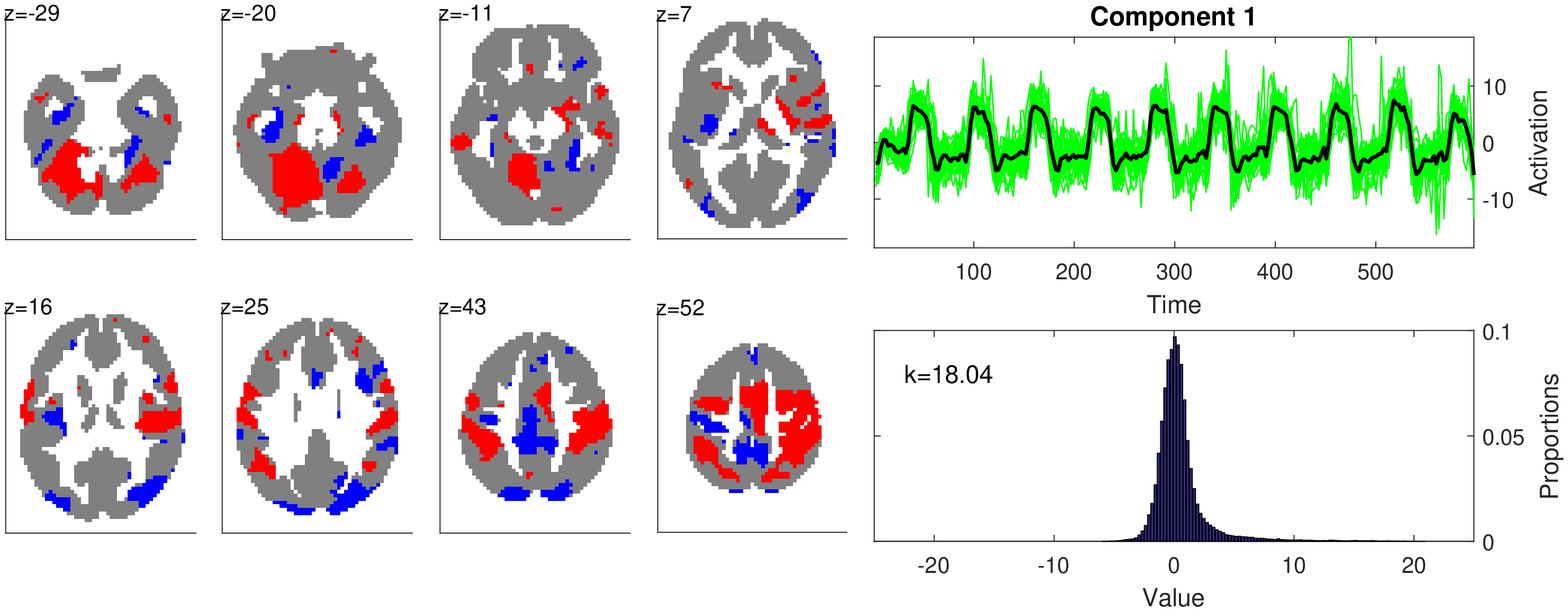} &
\includegraphics[width=.33\linewidth]{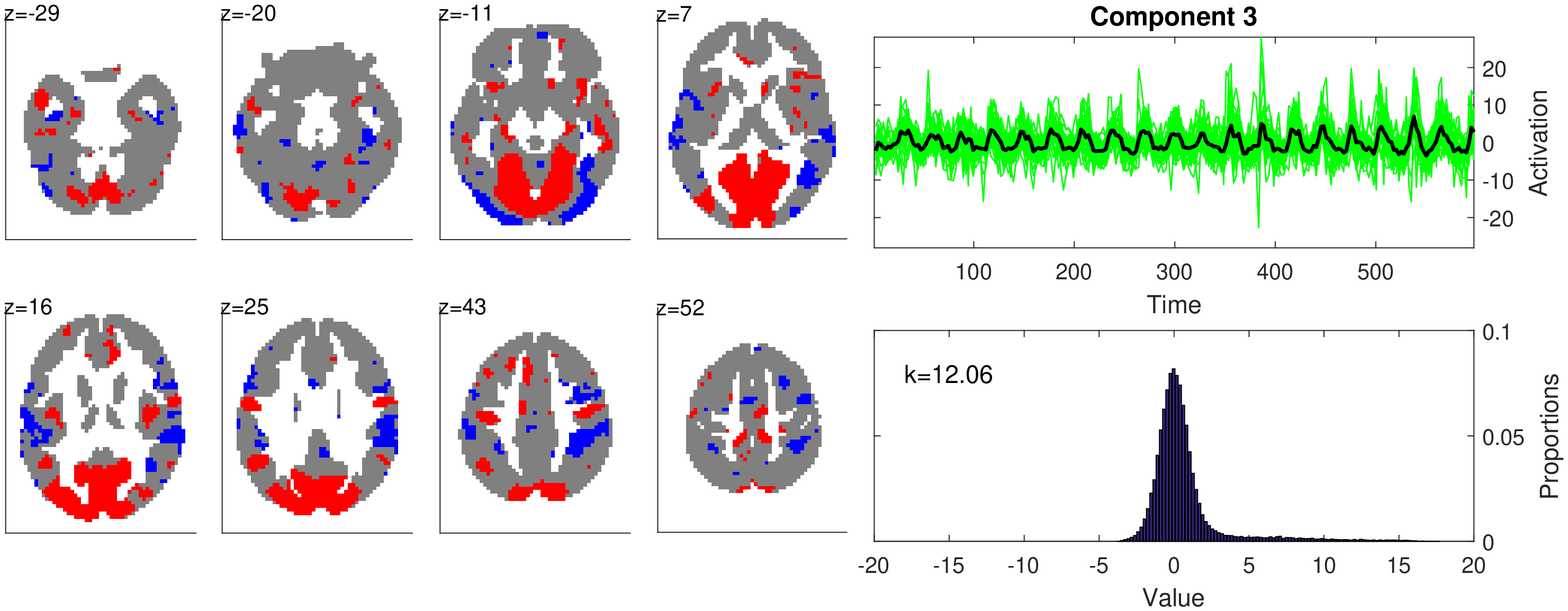}
\end{tabular}
\vspace{-.3cm}\caption{\textbf{MELODIC-ICA}: Left- and right sensorimotor cortex appear more lateralized, compared to psFA.} \label{fig:motor_ica}
\end{subfigure}
\caption{\textbf{Motor experiment} (psFA): The three estimated components for each model (psFA,pFA and MELODIC-ICA) which had highest correlation to sensorimotor (left and right) and visual related areas. For each component the following is shown; 1) eight slices with z-scored and thresholded ($>1$) spatial activation (red: positive, blue: negative). 2) histogram (100 bins) of the spatial elements. We also report the empirical kurtosis $k$. 3) temporal activation of the component (black = mean over subjects, green = individual subjects).}\label{fig:motor_fmri}
\end{figure*}

The Pearson correlation between the estimated components and a set of reference maps was then calculated. The reference maps were, the default mode network (DMN) from \cite{franco2009interrater} and eight anatomical regions from \cite{eickhoff2005new} which were: 1) Visual hOc1, hOc2, FG1, FG2; 2) Left, right sensoriomotor, left and right motor cortex. For each model a visual and two motor components with highest absolute Pearson correlation to the reference maps are shown in Fig.~\ref{fig:motor_fmri}. The components are sign corrected such that there is a positive correlation.  The components found by psFA and pICA , Fig.~\ref{fig:motor_sfa} and \ref{fig:motor_ica}, have more well-defined spatial and temporal activation than those found by pFA, Fig.~\ref{fig:motor_fa}. While pFA does capture the experimental design, the resulting spatial maps are more dense making them difficult to interpret. From the histograms, it is evident that both psFA and pICA enforce super-Gaussian distributions, which pFA does not.

\begin{figure*}
\centering
\begin{subfigure}[t]{.48\linewidth}
\centering
\includegraphics[width=.8\linewidth]{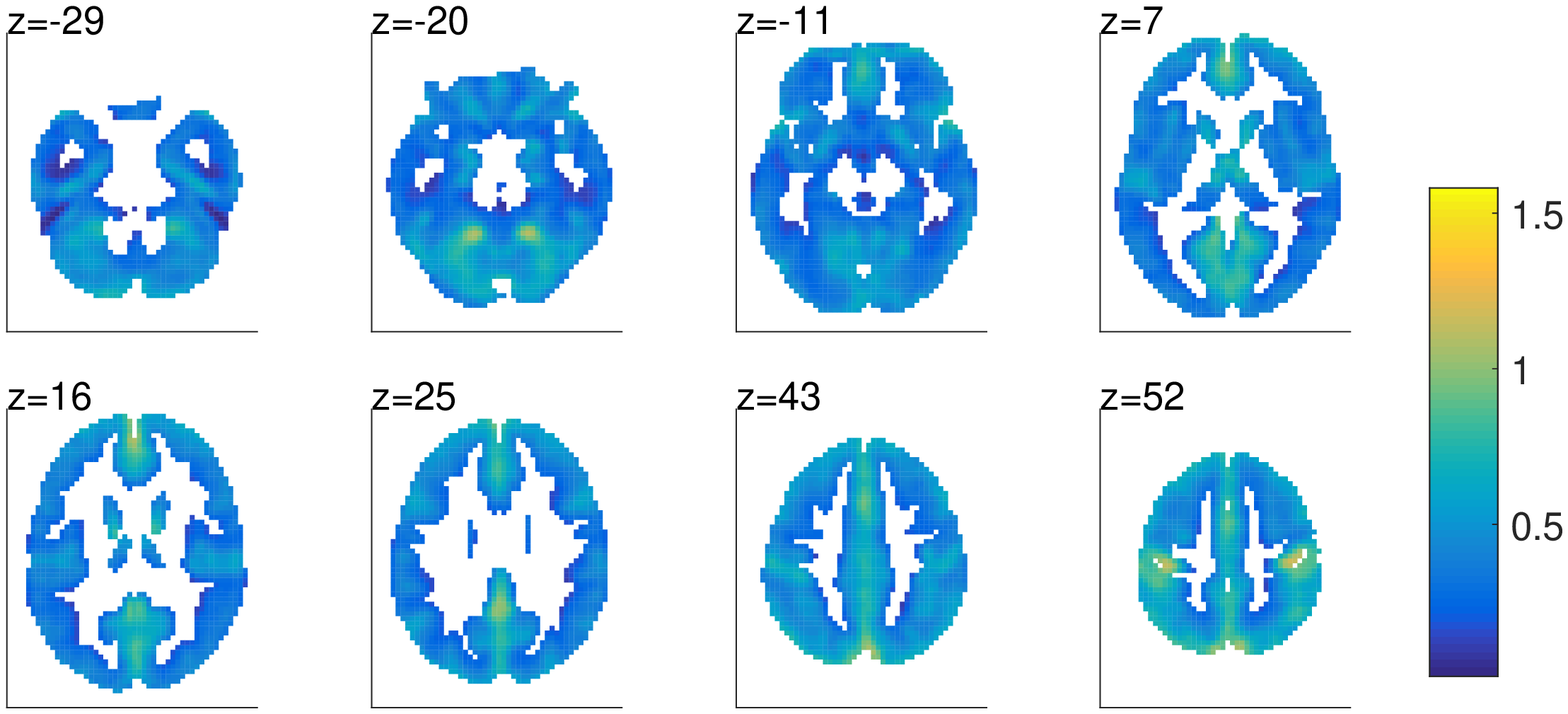}
\caption{\textbf{Motor experiment}: Regions showing high noise precision are related to the experimental design, i.e. motor cortex and visual area.}\label{fig:motor_fmri_noise}
\end{subfigure}
~\hfill~
\begin{subfigure}[t]{.48\linewidth}
\centering
\includegraphics[width=.8\linewidth]{img/noisemap_mean.eps}
\caption{\textbf{Resting state} (psFA): Regions with high noise precision are mostly located around the DMN.}\label{fig:rsfmri_noise}
\end{subfigure}
\caption{The expectation of log precision ($\<\log(\btau^{(b)}\>$), averaged over subjects/sessions.}
\end{figure*}

The expectation of the log precision of the noise $\btau$, averaged over subjects, is shown in Fig.~\ref{fig:motor_fmri_noise}. As the estimated precision varies over voxels, this hints that the assumption of heteroscedastic noise is supported by the data. Furthermore, the regions of high precision are related to the experimental design, where a high signal to noise ratio is expected. The noise precision estimates by pFA are similar those of psFA and are therefore not shown.

\begin{figure*}
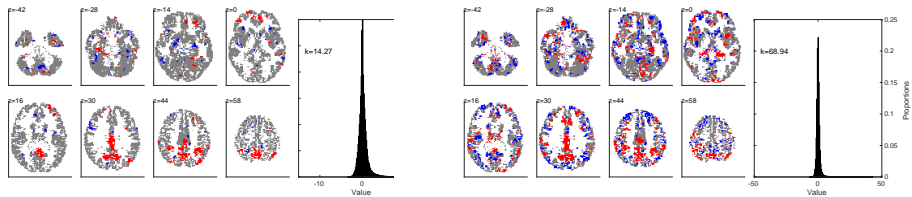

\centering
\begin{subfigure}[b]{.49\linewidth}
\includegraphics[width=\linewidth]{img/myconnect_MNI_D50_rep4_sameprior_d18.eps} 
\end{subfigure}
\begin{subfigure}[b]{.49\linewidth}
\includegraphics[width=\linewidth]{img/ICA_D85_d4.eps} 
\end{subfigure}
\caption{\textbf{Resting state experiment}: For psFA (left) and MELODIC-ICA (right) we show the component with highest spatial correlation to the default mode network. We show;  1) Eight z-scored and thresholded ($>1$) spatial activation slices  (red: positive, blue: negative). 2) A histogram (100 bins) of the spatial elements. We also report the empirical kurtosis $k$.}\label{fig:rsfmri}
\end{figure*}

\subsection{Resting state Experiment}
The psFA model was run for 2000 iterations with $D=50$ components and five random restarts. In Fig.~\ref{fig:rsfmri}, we show the component from psFA and pICA that has highest correlation with the DMN reference map (see sec.~\ref{sec:results_motor}). Both components  have super-Gaussian shape (cf. histograms and kurtosis), and seem to capture the posterior part of the DMN. However, it seems the psFA obtains a more sparse solution. Finally, we observe from the precision-noise maps in Fig.~\ref{fig:rsfmri_noise}, where it seems that highest certainty is found in areas contributing to DMN.

\vspace{-.4cm}\section{Summary}

In this work we investigated a scalable sparse probabilistic extension of factor analysis (psFA) for fMRI. We found in two data sets; a motor-task experiment with 29 subjects and a resting-state data set with 25 sessions, that inducing this form of sparsity results in ICA-like components. 
The probabilistic approach enables joint modeling of noise and quantification of parameters and their uncertainties. This comes at a computational cost, but due to the model structure a lot of computations are trivially parallelizable which we have exploited in the implementation. The fast inference scheme is currently limited by the memory on the single GPU used, but the proposed model easily extends to multiple GPUs. 
Future directions should be to investigate more advanced noise models and prediction on previously unseen data.


\bibliographystyle{unsrt}
\bibliography{latenvar}

\clearpage

\section*{Appendix}

\begin{center}
The derived details for probabilistic sparse factor analysis model for group level analysis are given. The model was first proposed by the authors and applied to functional magnetic resonance imaging\cite{hinrich2016psfa}. 
\end{center}

\subsection*{Probabilistic Sparse Factor Analysis}
The model's likelihood function, for a data array of size $V \times T \times B$ with all relevant parameters collected in $\btheta$, can be written as,
\begin{equation}
\mathcal{L}\left(\X|\btheta\right) = \prod_{b=1}^B\prod_{t=1}^T\DN\left(\x_t^{(b)}|\A\s_t^{(b)}+\bmu^{(b)},\diag\left(\btau^{(b)}\right)^{-1} \right),
\end{equation}
in which $\x_t^{(b)}$ is a vector of length $V$, $\A$ is a matrix of size $V\times D$, where D is the size of the latent space, $\s_t^{(b)}$ is a vector of size $D$ and $\btau^{(b)}$ is a vector of length $V$. 
The distributions of the parameters in the model are,
\begin{align*}
P(\A|\bm{\alpha}) =& \prod_{v=1}^V\DN \left(\a_v|\0,\diag\left(\balpha_{v}\right)^{-1}\right)\\
P(\S|\bgamma) =& \prod_{b=1}^B\prod_{t=1}^T\DN\left(\s_{t}^{(b)}|0,\diag(\bgamma)^{-1}\right)\\
P(\bmu) =& \prod_{b=1}^B \DN\left(\bmu^{(b)}|\0,\beta^{-1}\I_V\right)\\
P(\btau) =& \prod_{b=1}^B\prod_{v=1}^V\DGamma \left(\tau_{v}^{(b)}|a_\tau,b_{\tau_{v}^{(b)}}\right)\\
P(\bgamma) =& \prod_{d=1}^{D}\DGamma \left(\gamma_{d}|a_\gamma,b_{\gamma_{d}}\right) \\ 
P(\balpha) =& \prod_{v=1}^V\prod_{d=1}^D\DGamma \left(\alpha_{vd}|a_\alpha,b_{\alpha_{vd}}\right).
\end{align*}

Finding the parameters $\btheta$ from observed data $\X$ can be done by inferring the posterior distribution,
$$P(\btheta|\X) = \mathcal{L}(\X|\btheta)P(\btheta)/P(\X).$$
Unfortunately an exact inference is unfeasible for all but the simplest problems. Therefore an approximate solution is sought. While there are numerous ways to tackle this, we use variational Bayesian (VB) inference as used by \cite{Bishop1999-si} for VB principal components analysis. We use a mean field approximation, and find the following variational distribution to approximate the posterior.

\begin{align*}
Q(\A) =& \prod^{V}_{v=1} \DN(\a_v|\bmu_{\A_v},\bSigma_\A^{(v)})\\
Q(\S) =& \prod^{B,T}_{b,t=1}\DN(\s_t^{(b)}|\bmu_{\S_t}^{(b)},\bSigma_{\S}^{(b)})\\
Q(\bmu) =& \prod_{b=1}^B \DN\left(\bmu^{(b)}|\bmu_{\bmu}^{(b)},\bSigma_{\bmu}^{(b)}\right)\\
Q(\btau) =& \prod^{B,V}_{b,v=1}\DGamma(\tau_{v}^{(b)}|\tilde{a}_\tau,\tilde{b}_{\tau_{v}^{(b)}})\\
Q(\bgamma) =& \prod^{D}_{d=1}\DGamma(\gamma_{d}|\tilde{a}_\gamma,\tilde{b}_{\gamma_{d}})\\ 
Q(\balpha) =& \prod^{D,v}_{d,v=1}\DGamma(\alpha_{vd}|\tilde{a}_\alpha,\tilde{b}_{\alpha_{vd}})\\
\end{align*}

The moments of a distribution are then found by conditioning them on all other distributions and using free-form optimization (see \cite{Bishop1999-si}).

\subsection*{Update rules}
The found moments of the distributions are updated cyclically using Expectation-Maximization. In each iteration, after all distributions have been updated, the evidence lowerbound (ELBO) is calculated. After a number of iterations, when the relative change in ELBO is below a given threshold a set of local optimal parameters is identified.
\begin{align*}
\bSigma_\A^v =& \left(\diag\<\balpha_v\>+\sum_{b=1}^B\<\tau_v^{(b)}\>\<\S^{(b)}\S^{(b)^\top}\>\right)^{-1}\\
\bmu_\A^v =& \bSigma_\A^v \left( \sum_{b=1}^B\<\tau_v^{(b)}\>\sum_{t=1}^T\<\s_t^{(b)}\>\left(x_{tv}^{(b)}-\<\bmu_v^{(b)}\>\right) \right) \\
\bSigma_\S^{(b)} =& \left( \diag\<\bgamma\> + \<\A^\top \diag (\btau^{(b)}) \A \> \right)^{-1}\\
\bmu_{\S_t}^{(b)} =& \bSigma_\S^{(b)}\<\A^\top\>\diag\<\btau^{(b)}\>\left(\x_t^{(b)} - \<\bmu^{(b)}\>\right)\\
\bSigma_{\bmu}^{(b)} =& \left(\beta\I_V+\diag\<\btau^{(b)}\>\right)^{-1}\\
\bmu_{\bmu}^{(b)} =& \bSigma_{\bmu}^{(b)} \diag\<\btau^{(b)}\> \sum_{t=1}^T\left(\x_t^{(b)} - \<\A\>\<\s_t^{(b)}\>\right)\\
\tilde{a}_\alpha =& a_\alpha + \frac{1}{2} \quad, \quad \tilde{b}_{\alpha_{vd}} = b_{\alpha_{vd}}+\<a_{vd}^2\>\\
\tilde{a}_\gamma =& a_\gamma + \dfrac{1}{2}\sum_{b=1}^B\mathrm{T^{(b)}} \quad, \quad \tilde{b}_{\gamma_{d}} = b_{\gamma_{d}}+\dfrac{1}{2}\sum_{b=1}^B\tr\left(\<\s_d\s_d^\top\>\right)\\
\tilde{a}_{\tau^{(b)}} =& a_\tau + \frac{T^{(b)}}{2} \\ \tilde{b}_{\tau_{v}^{(b)}} &= b_{\tau_{v}^{(b)}}+\frac{1}{2}\Big[ ||\x_v^{(b)} ||^2_\mathrm{Fro} +  T^{(b)} \<\mu_v^{(b)^2}\> \\
&- 2 \left(\<\a_v\>\<\S^{(b)}\>+\<\mu_v^{(b)}\>\right)\x_v^\top \\
& + 2\<\a_v\>\<\S^{(b)}\>\1_{T^{(b)}}\<\mu_v^{(b)}\> \\
& + \tr\left(\<\a_v^\top \S^{(b)}\S^{(b)^\top}\a_v\>\right)\Big]
\end{align*}

Note $\<\cdot \>$ is the expected value under the variational distributions. Further, using the properties of the trace operator, the expected value of the expression in $\bSigma_\S^{(b)}$ and $\tilde{b}_{\tau_v^{(b)}}$ are determined to be,
\begin{eqnarray*}
\tr\left(\<\a_v^\top \S^{(b)}\S^{(b)^\top}\a_v\>\right) = \tr\left(\<\S^{(b)}\S^{(b)^\top}\>\bSigma_\A^v\right) \\ +\<\a_v^\top\>\<\S^{(b)}\S^{(b)^\top}\>\<\a_v\>
\end{eqnarray*}
and
\begin{eqnarray*}
\<\A^\top \diag (\btau^{(b)}) \A \> = \left(\sum_{v=1}^V \bSigma_A^v \<\tau_v^{(b)}\>\right)\\+\<\A^\top\> \diag \< \btau^{(b)}\> \< \A \> .
\end{eqnarray*}

\subsection*{Evidence Lowerbound (ELBO)}
The evidence lowerbound is the sum of all the expression in this section. It can be divided into to categories; 1) the expected value of the P-distributions under the Q-distributions (i.e. substituting the moments of the P-distributions for the moments of the Q-distributions). 2) The entropy of the Q-distributions.

For each P-distribution the expected value the corresponding Q-distribution is given below,
\begin{align*}
\< \log P(\A|\bm{\alpha})\> =& \sum_{v=1}^V -\frac{1}{2}\log(2\pi)+\frac{1}{2}\<\log \alpha_{vd}\>-\frac{1}{2}\<\alpha_{vd}\> \<a_{vd}^2\>\\
\< \log P(\S|\bgamma)\> =& \sum_{b=1}^B\sum_{t=1}^{T^{(b)}} -\frac{D}{2}\log(2\pi)+\left(\frac{1}{2}\sum_{d=1}^D\<\log \gamma_d\>\right)\\ &-\frac{1}{2}\tr\left(\diag\<\bgamma\>\<\s^{(b)}_t\s_t^{(b)^\top}\>\right)\\
\< \log P(\bmu)\> =& \sum_{b=1}^B -\frac{V}{2}\log(2\pi)+\frac{V}{2}\log(\beta)-\frac{1}{2}\beta \<\bmu_{\bmu}^{(b)^\top}\bmu_{\bmu}^{(b)}\>\\
\< \log P(\balpha)\> =& \sum_{d=1}^D\sum_{v=1}^V -\log (\Gamma(a_{\alpha}))+a_\alpha \log \left(b_{\alpha_{vd}}\right)\\ &+(a_\alpha-1)\<\log \alpha_{vd}\> - b_{\alpha_{vd}}\<\alpha_{vd}\> \\
\< \log P(\bgamma)\> =& \sum_{d=1}^{D} -\log (\Gamma(a_{\gamma}))+a_\gamma \log \left(b_{\gamma_{d}}\right)\\ &+(a_\gamma-1)\<\log \gamma_{d}\> - b_{\gamma_{d}}\<\gamma_{d}\>\\ 
\< \log P(\btau)\> =& \sum_{b=1}^B\sum_{v=1}^V -\log (\Gamma(a_{\tau}))+a_\tau \log \left(b_{\tau_{v}^{(b)}}\right)\\ &+(a_\tau-1)\<\log \tau_{v}^{(b)}\> - b_{\tau_{v}^{(b)}}\<\tau_{v}^{(b)}\>
\end{align*}

Finally, the entropy for each Q-distribution is,
\begin{align*}
-\< \log Q(\A)\> =& \sum_{v=1}^V \left[\frac{1}{2}\log |\bSigma_\A^v|+\frac{D}{2}(1+\log(2\pi))\right]\\
-\< \log Q(\S)\> =& \sum_{b=1}^B\sum_{t=1}^{T^{(b)}} \left[\frac{1}{2}\log |\bSigma_\S^{(b)}|+\frac{D}{2}(1+\log(2\pi))\right]\\
-\< \log Q(\bmu)\> =& \sum_{b=1}^B\left[\frac{1}{2}\log |\bSigma_{\bmu}^{(b)}|+\frac{V}{2}(1+\log(2\pi))\right]\\
-\< \log Q(\balpha)\> =& \sum_{d=1}^D\sum_{v=1}^V \log (\Gamma(\tilde{a}_{\alpha_{vd}}))-(\tilde{a}_\alpha-1)\psi(\tilde{a}_\alpha) \\ &-\log(\tilde{b}_{\alpha_{vd}})+\tilde{a}_\alpha \\
-\< \log Q(\bgamma)\> =& \sum_{d=1}^{D} \log (\Gamma(\tilde{a}_\gamma))-(\tilde{a}_\gamma-1)\psi(\tilde{a}_\gamma)-\log(\tilde{b}_{\gamma_d})+\tilde{a}_\gamma\\ 
-\< \log Q(\btau)\> =& \sum_{b=1}^B\sum_{v=1}^V \log (\Gamma(\tilde{a}_{\tau^{(b)}}))-(\tilde{a}_{\tau^{(b)}}-1)\psi(\tilde{a}_\tau) \\ &-\log(\tilde{b}_{\tau^{(b)}_v})+\tilde{a}_{\tau^{(b)}}
\end{align*}

\subsection*{Implementation}
A MATLAB implementation is available\footnote{\url{https://brainconnectivity.compute.dtu.dk/}}. The implementation is limited to the use of a single GPU card, as well as analysis with $T^{(i)}=T^{(j)}, \forall i,j$. These limitations were deemed acceptable for the work in \cite{hinrich2016psfa}, as subjects with differing timesteps are not widespread in the field.

\end{document}